\documentclass{aa}
\usepackage{graphicx}
\usepackage{txfonts}

\newcommand{\sax}    {\object{SAX J2103.5+4545}}

\def\simless{\mathbin{\lower 3pt\hbox
     {$\rlap{\raise 5pt\hbox{$\char'074$}}\mathchar"7218$}}}   
\def\simmore{\mathbin{\lower 3pt\hbox
     {$\rlap{\raise 5pt\hbox{$\char'076$}}\mathchar"7218$}}}   

\def\msun{~{\rm M}_\odot}
\def\rsun{~{\rm R}_\odot}

\begin{document}


\title{Discovery of the optical counterpart to the X-ray pulsar  \\ SAX J2103.5+4545}

\subtitle{}

\author{P. Reig\inst{1}
\and I. Negueruela\inst{2}
\and J. Fabregat\inst{3}
\and R. Chato\inst{1}
\and P. Blay\inst{1}
\and F. Mavromatakis\inst{4}
}

\institute{
G.A.C.E, Instituto de Ciencias de los Materiales, Universitat de Valencia, 46071 Paterna-Valencia, Spain\\
\email{pablo.reig@uv.es}
\email{pere.blay@uv.es}
\email{rachid.chato@uv.es}
\and Departamento de F\'{\i}sica, Ingenier\'{\i}a de Sistemas y Teor\'{\i}a
de la Se\~nal, Universidad de Alicante, E-03080 Alicante, Spain\\
\email{ignacio@dfists.ua.es}
\and Observatorio Astron\'omico de Valencia, Universitat de Valencia, 46071 Paterna-Valencia, Spain\\
\email{juan.fabregat@uv.es}
\and University of Crete, Physics Department, PO Box 2208, 710 03
Heraklion, Crete, Greece \\
\email{fotis@physics.uoc.gr}
}

\authorrunning{Reig et~al.}
\titlerunning{The optical counterpart to SAX\,~J2103.5+4545}

\offprints{P. Reig, \\ \email{pablo.reig@uv.es}}

\date{Received / Accepted }

\abstract{

We report optical and infrared photometric and spectroscopic observations
that identify the counterpart to the 358.6-s X~-~ray transient pulsar \sax\
with a moderately reddened V=14.2 B0Ve star. This identification makes
\sax\ the Be/X-ray binary with the shortest orbital period known, $P_{\rm
orb}= 12.7$ days. The amount of absorption to the system has been
estimated to be $A_V=4.2\pm0.3$, which for such an early-type star implies
a distance of about 6.5 kpc. The optical spectra reveal major and rapid
changes in the strength and shape of the H$\alpha$ line. The H$\alpha$
line was initially observed as a double peak profile with the ratio of the
intensities of the blue over the red peak greater than one ($V/R > 1$).
Two weeks later this ratio reversed ($V/R< 1$).  Subsequently, in less
than a month, the emission ceased and H$\alpha$ appeared in absorption.
This fast spectral variability is interpreted within the viscous decretion
disc model  and demonstrates the significant role of the neutron star on
the evolution of the circumstellar disc around the Be star.  The
implications of the small orbit and moderate eccentricity  on the spin
period of the neutron star are discussed.

}

\maketitle

\section{Introduction} \label{introduction}

The X-ray transient \sax\ was discovered by the BeppoSAX Wide Field Camera
(WFC) during an outburst in February 1997 (Hulleman et al. \cite{hul98}).
The source was active for about eight months and reached a peak intensity
of 20 mCrab (2--25 keV) on April 11, 1997. The X-ray flux showed pulsed
emission with a period of 358.61 s. An absorbed power law  ($dN/dE\propto
e^{-N_H \sigma(E)}E^{-\Gamma}$) with photon number index $\Gamma=1.27$ and
column density $N_H=3.1 \times 10^{22} {\rm cm^{-2}}$ was used to fit the
X-ray energy spectrum.  

A second outburst was detected on October 25, 1999 by the All Sky Monitor
on board RXTE (Baykal et al. \cite{bay00}), reaching a peak intensity of 27
mCrab (2-12 keV) three days later. The longer duration of this activity
period (about 14 months) and regular monitoring allowed a Doppler shifts
analysis of the pulsations resulting in the derivation of the orbital
parameters (Baykal et al. \cite{bay00}): the system has a moderately
eccentric orbit ($e=0.4\pm0.2$) and an orbital period of 12.68$\pm$0.25
days. The RXTE 3--50 keV X-ray energy spectrum was well described by an
absorbed power law, modified by a cutoff at around 8 keV. In addition, the
fits required an emission iron line at 6.4 keV. The power-law index and the
flux of the iron line were seen to vary with X-ray intensity (Baykal et al.
\cite{bay02}).

Inam et al. (\cite{ina04}) reported the detection of a 0.044 Hz QPO using
XMM-Newton data, and estimated the magnetic field of the neutron star to
be $\sim 7\times 10^{12}$ G. These authors also found that the energy
spectrum at low energies required a soft component with a characteristic
temperature of 1.9 keV (if modeled as a blackbody).

The nature of the optical counterpart to \sax\ was uncertain.  The shape
of the X-ray energy spectrum, the transient behaviour and the increase of
the X-ray intensity around periastron passages (Baykal et al.
\cite{bay02}) point toward a Be/X-ray binary. However, \sax\ does not
follow the correlation between the orbital period and the spin period 
(Corbet \cite{cor86}) of other Be/X-ray systems. Although Hulleman et al.
(\cite{hul98}) tentatively suggested that the star HD~200709 might be
linked to \sax, its position outside the BeppoSAX error box
($\sim2^{\prime}$) and its late spectral type (B8V) posed a serious
hindrance to its candidacy as the correct optical counterpart.  


In this work we present ground-based observations that identify the
optical counterpart to \sax\ and classify it as a Be/X-ray binary. The fast
optical spectral variability, X-ray behaviour and position in the $P_{\rm
spin}-P_{\rm orb}$ diagram are explained within the viscous decretion
disc model.

\begin{figure}
\resizebox{\hsize}{!}{\includegraphics{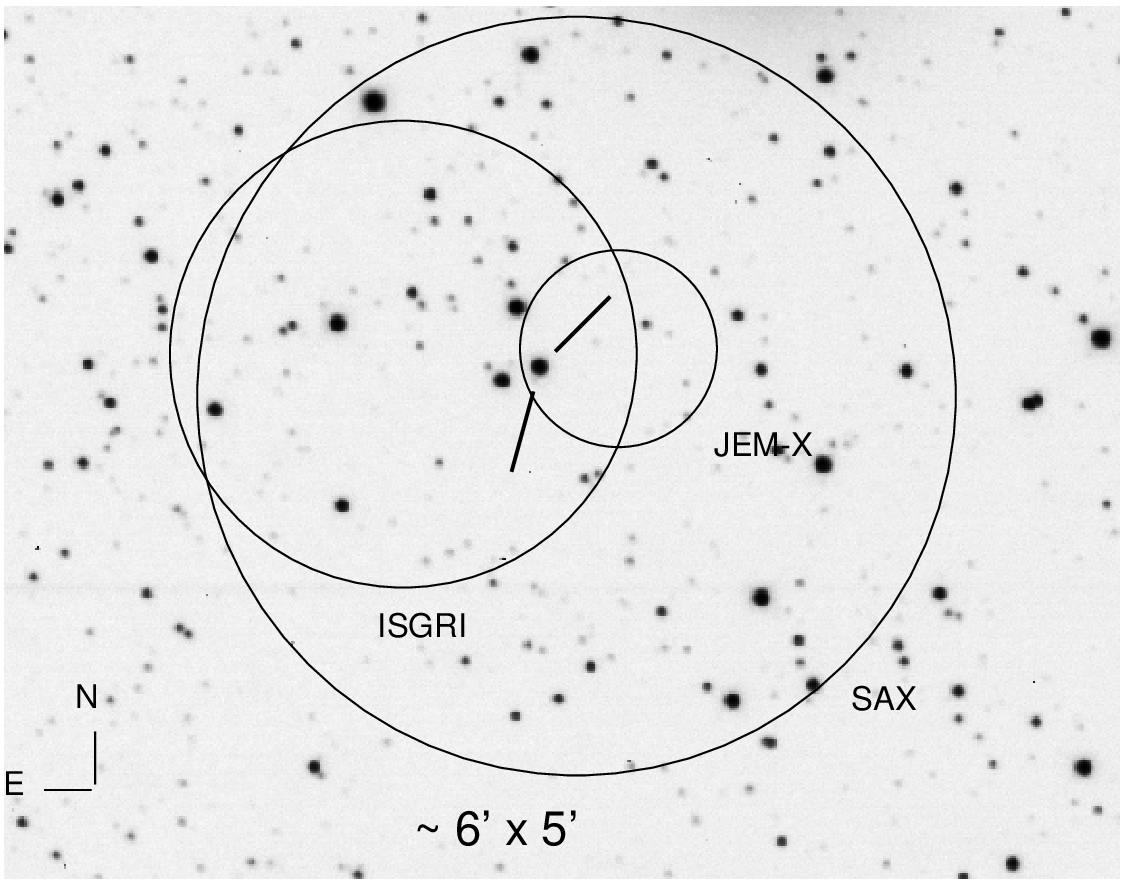}}
\caption[]{Optical $V$ filter image of the field around \sax, with the
X-ray position uncertainty circles from the WFC onboard BeppoSAX 
(99\% confidence level) and ISGRI and JEM-X onboard
INTEGRAL (90\% confidence level). The coordinates 
of the proposed optical counterpart are RA=21:03:35.7 DEC=+45:45:04 (Eq. 2000)}
\label{Vimage}
\end{figure}

\section{Observations} \label{observations}

\subsection{Optical observations}

The field around the best-fit X-ray position given by the BeppoSAX Wide
Field Camera was observed through the Johnson $B$, $V$, and $R$ filters
from the 1.3\,m telescope of the Skinakas observatory on June 8, 2003 and
August 24, 2003. using a $1024 \times 1024$ SITe CCD chip with a 24 $\mu$m
pixel size (corresponding to $0.5^{\prime\prime}$ on sky). On June 8, 2003
the field was also observed through an interference filter centred at 6563
\AA, and a width of 10 \AA\ (H$\alpha$ filter). Standard stars from the
Landolt list (Landolt \cite{lan92}) were used for the transformation
equations.  Reduction of the data was carried out in the standard way
using the IRAF tools for aperture photometry. The photometric magnitudes
are given in Table~\ref{phot}. Figure~\ref{Vimage} shows a V-band image of
the field of \sax. The $\sim 2^{\prime}$ BeppoSAX WFC (Hulleman et al.
\cite{hul98}), $\sim 1.2^{\prime}$ INTEGRAL ISGRI and $\sim
30^{\prime\prime}$ INTEGRAL JEM-X (Blay et al. \cite{bla04}) error
circles are also shown.

Optical spectroscopic observations were obtained from the  Skinakas
observatory in Crete (Greece) on August 1, 2003 and October 6 and 8, 2003
and from the observatory of  El Roque de los Muchachos in La Palma (Spain)
on the night August 17, 2003. The 1.3\,m telescope of the Skinakas
Observatory was equipped with a 2000$\times$800 ISA SITe CCD and a 1302
l~mm$^{-1}$ grating, giving a nominal dispersion of $\sim$1 \AA/pixel.
The  William Herschel 4.2\,m telescope (WHT) was equipped with the ISIS
double-armed (R1200B and R1200R gratings) spectrograph, giving a
dispersion of 0.23 \AA/pixel. The blue (3900--4700 \AA) and red
(6100--6900 \AA) spectra were obtained with the  4096$\times$2048 (13.5
$\mu$m) pixels EEV12 CCD and MARCONI2 CCD, respectively. Further spectra
were obtained through the service time scheme from the WHT on the night of
September 14, 2003. The instrumental set-up was the same as above except
for the fact that lower resolution gratings (R600B and R600R) were
employed, providing a dispersion of 0.44 \AA/pixel but a larger wavelength
coverage. 

The reduction of the spectra was made using the STARLINK {\em Figaro}
package (Shortridge et~al. \cite{sho01}), while their analysis was
performed using the STARLINK {\em Dipso} package (Howarth et~al.
\cite{how98}).

\begin{figure}
\resizebox{\hsize}{!}{\includegraphics{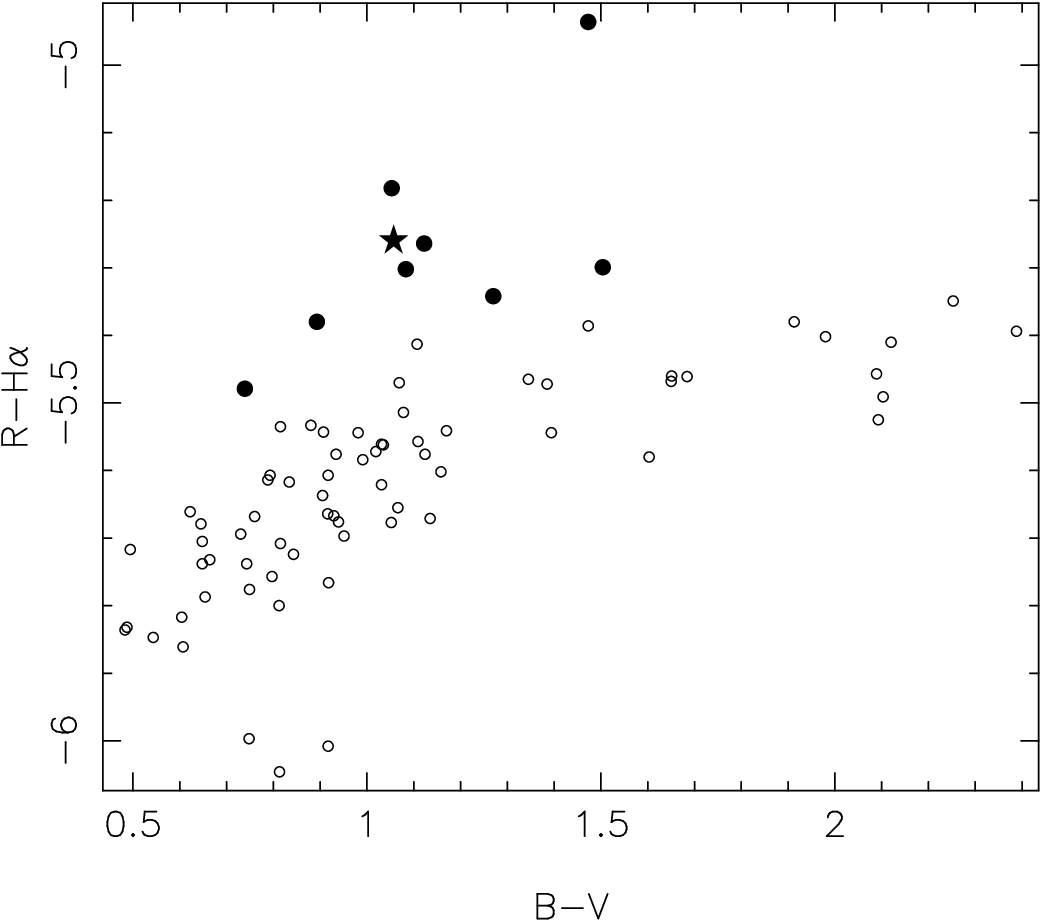}}
\caption[]{Colour-colour diagram of the field around \sax. The star
symbol marks the position of the proposed optical counterpart. 
Filled-circles represent strong H$\alpha$ emitters, while open circles 
correspond to field stars.}
\label{cd}
\end{figure}

\subsection{Infrared observations}

Infrared photometry in the JHK bands was done at the 1.5m. Carlos 
S\'anchez Telescope (TCS) located at the Teide Observatory in Tenerife, 
Spain. The instrument used was the Cain-II camera, equiped with a 
256$\times$256 HgCdTe (NICMOS 3) detector. Instrumental magnitudes were 
obtained from the images by means of the IRAF tools for aperture 
photometry. Instrumental values were transformed to the standard system 
defined by the standard star list published by Hunt et al. (\cite
{hun98}). The  accuracy of the standard values, defined as the standard
deviation of the  mean catalogue minus transformed values for the standard
stars, is 0.03  mag. in all three bands. The obtained values are given in
Table~\ref{phot}. 
An important variation of about 0.5 mag. in all IR bands from 24 to 27 
August is noticeable. This is in contrast with the small variation from 27 
August to 05 December, and also with the small variations in the optical 
bands. To check the reliability of the data we have obtained 
photometry of seven nearby stars present in the same images. All of them 
appear constant in all bands, within the accuracy of the photometry. Hence 
we can safely exclude the possibility of any instrumental effect, and 
confirm that in August 24 SAX J2103.5+4545 was significantly fainter in 
all IR bands.     

\begin{table}
\begin{center}
\caption{Optical and infrared magnitudes of \sax.}
\label{phot}
\begin{tabular}{lcccc}
\hline \hline \noalign{\smallskip}
\multicolumn{5}{c}{Optical} \\
\noalign{\smallskip} \hline \noalign{\smallskip}
Date		&B		&V		&R	&I  \\
\noalign{\smallskip} \hline \noalign{\smallskip}
08/06/03	&15.34$\pm$0.02	&14.22$\pm$0.02	&13.48$\pm$0.02 &-\\
24/08/03&15.36$\pm$0.03	&14.27$\pm$0.02 &13.59$\pm$0.03 &12.87$\pm$0.03 \\
\noalign{\smallskip} \hline \noalign{\smallskip}
\multicolumn{5}{c}{Infrared} \\
\noalign{\smallskip} \hline \noalign{\smallskip}
Date		&J		&H		&K	&\\
\noalign{\smallskip} \hline \noalign{\smallskip}
24/08/03 & 11.97$\pm$0.03 & 11.42$\pm$0.03 & 11.20$\pm$0.03 \\
27/08/03 & 11.38$\pm$0.03 & 10.90$\pm$0.03 & 10.67$\pm$0.03 \\
05/12/03 & 11.41$\pm$0.03 & 10.97$\pm$0.03 & 10.75$\pm$0.03 \\
04/01/04 & 11.40$\pm$0.04 & 10.96$\pm$0.05 & 10.75$\pm$0.04 \\
07/01/04 & 11.40$\pm$0.04 & 10.94$\pm$0.04 & 10.71$\pm$0.03 \\
\noalign{\smallskip} \hline
\end{tabular}
\end{center}
\end{table}

\section{Results}

\subsection{The colour-colour diagram}

The instrumental magnitudes corresponding to the $B$, $V$, $R$ and
H$\alpha$ filters obtained during the June 8, 2003 observations were used
to define a "blue" colour ($B-V$) and a "red" colour (R--H$\alpha$). Then
a colour-colour diagram was constructed by plotting the red colour as a
function of the blue colour as shown in Fig.\ref{cd}.  Stars with a
moderately or large H$\alpha$ excess can be distinguished from the rest
because they deviate from the general trend and occupy the upper parts of
the diagram (filled circles in Fig.\ref{cd}). This kind of diagrams have
been successfully used to identify optical counterparts in the Magellanic
Clouds (Grebel \cite{gre97}; Stevens et al. \cite{ste99}). Of the
approximately 80 stars that  were analysed in the $\sim 8^{\prime} \times
8^{\prime}$ field of view of the 1.3\,m telescope about 8 were bright
enough in H$\alpha$ to call our attention. The position of each one of
these potential candidates with respect to the satellites error circle was
then checked. Of these relatively strong H$\alpha$ emitters only one lied
inside the WFC BeppoSAX satellite position uncertainty circles
(Fig.\ref{Vimage}).  As well as appearing bright in H$\alpha$, this star
is highly reddened, $(B-V)>1$, making a good case for a Be classification.
However, the photometric data are not conclusive since other type of
systems, such as supergiants or late-type giants can also show emission
lines and/or have a very reddened spectrum. The final confirmation that
the system contains an early-type star was provided by the spectroscopic
observations, which also revealed H$\alpha$ in emission (Fig.~\ref{ha}).
Even more significant than the emission itself it is the double-peak
profile, since it can be interpreted as coming from the circumstellar
envelope of a Be star.

\subsection{Spectral type}

\begin{figure*}
\begin{center}
\includegraphics[width=16cm,height=9cm]{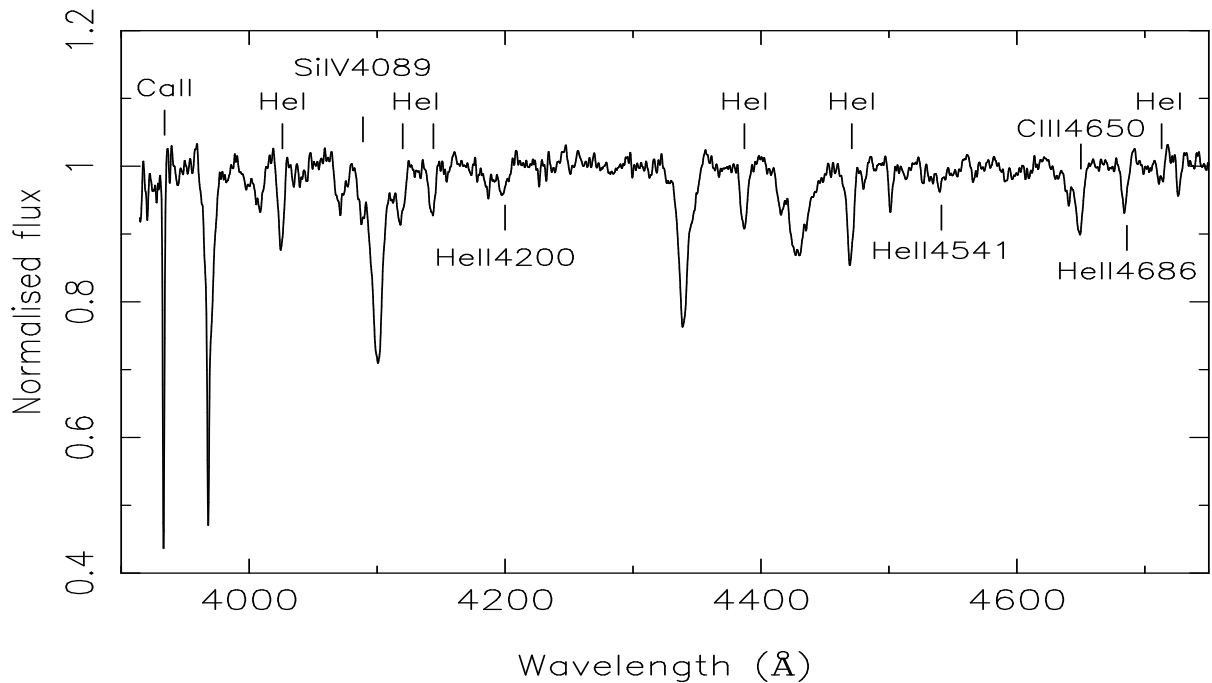}
\caption[]{Blue spectrum showing the lines used for spectral
classification. The spectrum was taken from the WHT on August 17, 2003. The
He I lines marked are $\lambda$4026, $\lambda$4121, $\lambda$4144, $\lambda$4387, 
$\lambda$4471, $\lambda$4713. Also shown is the insterstellar line Ca\,II
$\lambda$3934.}
\label{sp}
\end{center}
\end{figure*}

A visual comparison of the \sax\ spectrum (Fig.~\ref{sp}) with those of MK
standards in the Walborn \& Fitzpatrick (\cite{wal90}) atlas reveals that
the ratios He\,I $\lambda$4026/He\,II $\lambda$ 4200 and He\,I
$\lambda$4713/He\,II $\lambda$4686 are very similar to those of the
standard star $\upsilon$ Ori (B0V). Indeed, the presence of He\,II
($\lambda$4200,  $\lambda$4686) indicates that the spectral type of the
optical star is earlier than B1. Specifically, He\,II $\lambda$4686 is
last seen at B0.5-B0.7. On the other hand, the weakness of He\,II
$\lambda$4541 relative to He\,I $\lambda$4471 indicates an spectral type
later than O9.

With regard to the luminosity classification, the weakness of Si\,III 
$\lambda$4552--68 and the fact that the line intensitie of He\,II
$\lambda$4686 is larger than that of He\,I $\lambda$4713 seem to indicate
a main-sequence star. Other luminosity indicators are the ratios of He
lines to nearby metallic lines. \sax\ shows He\,II $\lambda$4686/CIII
$\lambda$4650 $\simless$ 1 and He\,I $\lambda$4144/Si\,IV 4089 $\sim$ 1. A
main-sequence star would have those ratios $\sim$1 and $\simmore$ 1,
respectively. Therefore, we conclude that the optical counterpart to the
X-ray accreting pulsar \sax\ is a B0V star, although a subgiant cannot be
excluded. This classification puts \sax\ in the group of Be/X-ray
binaries.

\subsection{Evolution of H$\alpha$}

Profile changes in emission lines, particularly, H$\alpha$, can be used to
trace the dynamical evolution of the Be envelope (Negueruela et al. 2001).
The H$\alpha$ profile obtained from each observing run is presented in
Fig.\ref{ha}. The spectra revealed a highly variable H$\alpha$ emission
line. Table~\ref{ewha} gives the measurements of the equivalent width of
the H$\alpha$ line for each observing run. Negative values indicate that
the line is in emission. In about two weeks  the ratio of the blue to the
red peak reversed and  the strength of the H$\alpha$ line decreased by
$\sim$ 50\%. The peak separation, however,  remained constant at about 440
km s$^{-1}$. The line profile of the August 17, 2003 spectrum is
reminiscent of the "shell class", i.e., a double peak emission with
central absorption presumably due to self-absorption as a consequence of a
high inclination angle. Subsequently, in less than a month, the line
appears in absorption. A very small amount of remaining emission might
still be present in the September 14, 2003 spectrum. For a normal
absorption-line B0V star the H$\alpha$ equivalent width is approximately
+3.5 \AA\ (Table 9.1 in Jaschek \& Jaschek \cite{jas87}). The lastest
spectra seem to indicate that a new emission phase has begun. The October
8, 2003 spectrum shows incipient emission at the bottom of the absorption
profile. 

\begin{table}
\begin{center}
\caption{H$\alpha$ equivalent width measurements. Errors are $\le$ 10\%}
\label{ewha}
\begin{tabular}{cccc}
\noalign{\smallskip} \hline \noalign{\smallskip}
Date	&MJD	&EW(H$\alpha$)	&Telescope \\
	&	&(\AA)		&	\\
\hline 
01/08/03&52853	&--2.20	&SKI	\\
17/08/03&52869	&--1.07	&WHT	\\
14/09/03&52897	&+2.32	&WHT	\\
06/10/03&52919	&+2.15	&SKI	\\
08/10/03&52921	&+1.95	&SKI	\\
\hline 
\end{tabular}
\end{center}
\end{table}

\begin{table}
\begin{center}
\caption{Diffuse interstellar lines used to estimate the reddening to \sax.}
\label{is}
\begin{tabular}{cccc}
\hline \hline \noalign{\smallskip}
Line	&EW$^a$ (m\AA)	&EW (m\AA)	&$E(B-V)$ \\
	&14/09/03	&17/08/03	& \\
\noalign{\smallskip} \hline \noalign{\smallskip}
6010	&213		&-		&-  \\
6195	&76		&100		&1.38$\pm$0.06 \\
6202	&341		&300		&1.42$\pm$0.18 \\
6269	&202		&215		&1.27$\pm$0.21 \\
6376/79	&193		&215		&1.24$\pm$0.07 \\
6613	&310		&318		&1.49$\pm$0.13 \\
\hline
Average	&		&		&1.36$\pm$0.10 \\
\noalign{\smallskip} \hline
\multicolumn{4}{l}{a: average of six measurements}
\end{tabular}
\end{center}
\end{table}

\subsection{Reddening and distance}

\begin{figure}
\resizebox{\hsize}{!}{\includegraphics{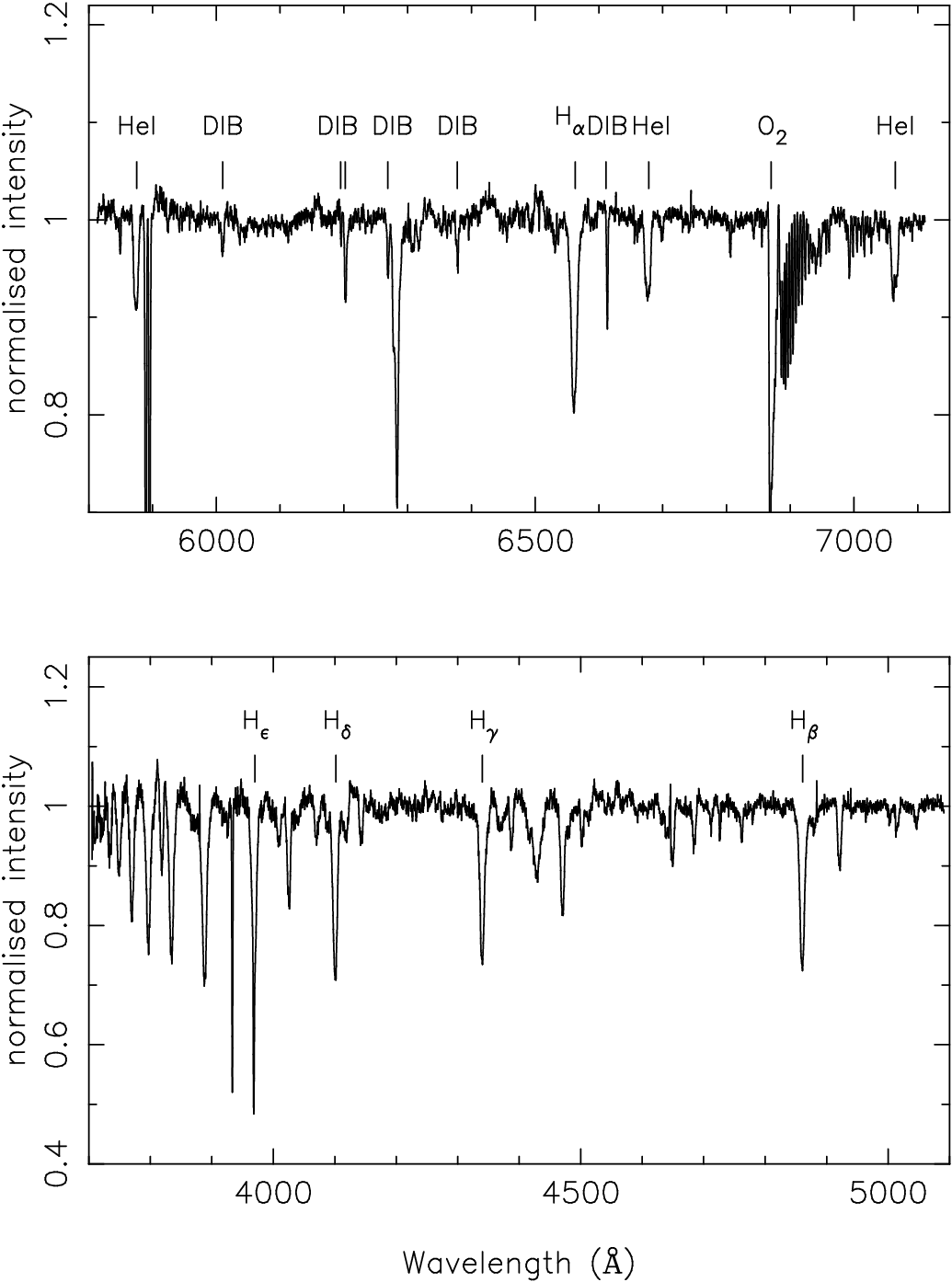}}
\caption[]{Blue- and red-end spectra of the optical companion of \sax. We
have indicated the diffuse interstellar bands used for the derivation of
the reddening and the Helium lines $\lambda$5875, $\lambda$6678,
$\lambda$7065. The Balmer series up to $H_{13}$ can be seen. 
The spectra were taken from the WHT on September 14, 2003}
\label{redblue}
\end{figure}

Besides the H$\alpha$ and He\,I $\lambda6678$ lines, the red spectrum
(Fig.\ref{redblue}) contains several strong diffuse interstellar bands
(DIB), which can be used to estimate the amount of interstellar absorption
toward the source (Herbig \cite{her75}; Herbig \& Leka \cite{her91};
Galazutdinov et al. \cite{gal00}). Table~\ref{is} summarises the
measurements of the equivalent width and the corresponding estimated value
of the colour excess $E(B-V)$ according to the linear relations of Herbig
(\cite{her75}) for the WHT observations. The quoted values for the
September 14, 2003 spectra are the average of six measurements. Those of 
August 17, 2003 correspond to one spectrum, with higher spectral
resolution.  The colour excess estimated from the DIBs is
$E(B-V)=1.36\pm0.10$, where the error is the standard deviation of the
averaged values. This colour excess agrees with that estimated from the
photometric data. A B0V star has an intrinsic colour $(B-V)_0=-0.27$
(Wegner \cite{weg94}) and taking the measured photometric colour
$(B-V)=1.10$ we derive and excess $E(B-V)=1.37$.  Likewise,  assuming
the interstellar extinction law $E(H-K)=0.20E(B-V)$ and the intrinsic
colour $(H-K)_0=-0.05$ for a main-sequence B0 star (Koornneef
\cite{koo83}) we obtain $E(B-V)=1.35$.

Finally, the distance can be estimated from the distance-modulus relation.
Taking the standard law $A_V=3.1 E(B-V)$ and assuming an
average absolute magnitude for a B0V star of $M_V=-4.2$ (Vacca et al.
\cite{vac96}) the distance to \sax\ is estimated to be $\sim$ 6.5$\pm$0.9
kpc.  This error includes the errors in $m_V$ (0.02) and $A_V$ (0.3),
but assumes no error in the absolute magnitude $M_V$.

\subsection{Rotational velocity}

Be stars are fast rotators. They have, on average, larger observed
rotational velocities than B stars as a group (Slettebak \cite{sle82}).
The determination of the rotational velocity is important because it has
been seen to correlate with emission characteristics (Briot \cite{bri86};
Mennickent et al. \cite{men94}). The rotational velocity can be estimated
by measuring the full width at half maximum of He I lines (e.g. Steele et
al. \cite{stee99}). We obtain $v \sin i=240\pm20$ km s$^{-1}$, which
compares to the value of 246$\pm$16 km s$^{-1}$ given for weak-emission
early-type shell stars (Mennickent et al. \cite{men94}). As a comparison
other rotational velocities in Be/X-ray binaries are:  $v \sin i=200\pm30$
km s$^{-1}$ in LS I +61 235/RX J0146.9+6121 (Reig et al. \cite{reig97}),
$v \sin i=290\pm50$ km s$^{-1}$ in V635 Cas/4U 0115+63 (Negueruela \&
Okazaki \cite{negoka01}), $v \sin i=240\pm20$ km s$^{-1}$ in  LS 992/RX
J0812.4--3114 (Reig et al. \cite{reig01}).

\begin{figure}
\resizebox{\hsize}{!}{\includegraphics{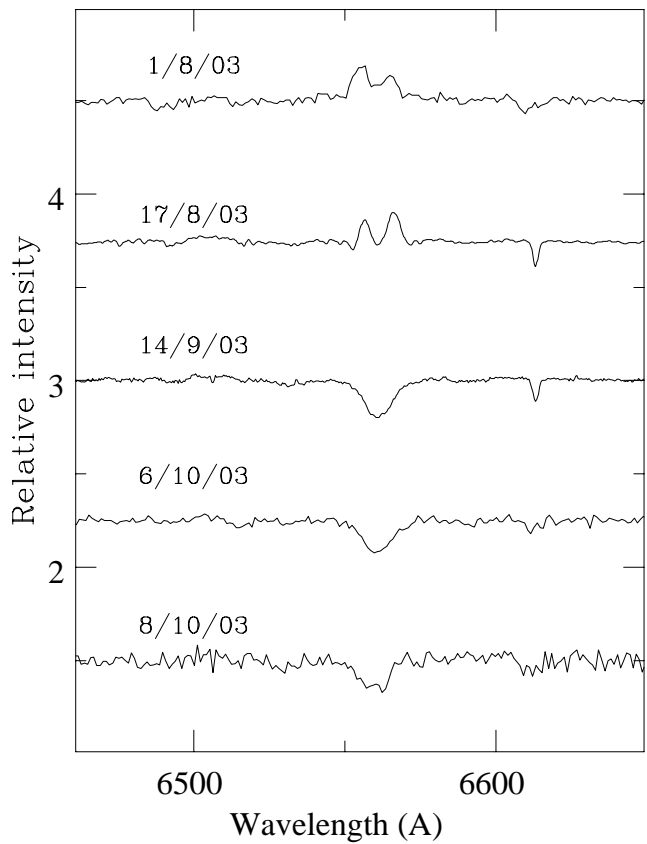}}
\caption[]{Evolution of the H$\alpha$ line.}
\label{ha}
\end{figure}

\section{Discussion}

\subsection{\sax: a new Be/X-ray binary}

The two most significant observational characteristics of Be stars in the
optical/infrared band are the emission lines and the infrared excess,
compared with non line-emitting B stars of the same spectral type. These
characteristics are associated with an equatorial outflow of material
expelled from the rapidly rotating Be star that forms a quasi-Keplerian
disc, whose effective temperature is 0.5--0.8 cooler than that of the
photosphere (Millar \& Marlborough \cite{mil99}). Consequently, Be stars
tend to appear redder than indicated by their spectral type. The slightly
bluer colours of the source measured in August with respect to the June
run and the progressive decrease of the  H$\alpha$ line intensity agrees
with a weakening of the disc. Likewise, the rapid and large amplitude of
variability of the infrared and H$\alpha$ intensities indicate a perturbed
disc. Nevertheless, the excellent agreement between the colour excess
derived from the photometry and from the strength of
interstellar lines (which are not affected by the circumstellar emission)
indicates that the amount of material in the disc must be scarce during
our observations. 

The absorption derived from the X-ray data is, however, considerably
higher than that derived from the optical observations. The value of the
hydrogen column density measured by the X-ray missions ranges between
$(3.1-3.8) \times 10^{22}$ cm$^{-2}$ (Hulleman et al. \cite{hul98}; Baykal
et al. \cite{bay02}), which represents $E(B-V)\approx 6$ (Predehl \&
Schmitt \cite{pre95}) or $E(B-V)\approx 5$ (Ryter et~al. \cite{ryt75};
Gorenstein \cite{gor75}), assuming $E(B-V)=A_V/3.1$ (Rieke \& Lebofsky
\cite{rie85}).  This discrepancy can be explained by the different
activity state of the source during the X-ray and optical observations.
While the X-ray observations of Hulleman et al. (\cite{hul98}) and Baykal
et al. (\cite{bay02}) took place during outbursts, the optical
observations were carried out when the source was weakly X-ray active ---
according to contemporaneous RXTE/ASM measurements (see also Stark et al.
\cite{sta03} and Blay et al. \cite{bla04}). In the Be/X-ray paradigm the
high-energy radiation is consequence of accretion from the circumstellar
disc. Thus during an outburst large amounts of material from the optical
companion are expected. The detection of an iron line at $\sim$ 6.4 keV
supports the presence of circumstellar material. In contrast, no or small
amounts of matter are accreted during low-activity X-ray states.

\sax\ displays the so-called type I outbursts, i.e., recurrent and
moderate ($L_{\rm X} \simless 10^{37}$ erg s$^{-1}$) increases of X-ray
intensity covering a small fraction of the orbit and modulated with the
orbital period. This transient behaviour is typical of Be/X-ray binaries
and has been reported in a number of systems (e.g. Bildsten et al.
\cite{bil97}). Considering the distance estimated above, 6.5 kpc, the
X-ray luminosity at the peak of the outbursts is $\sim 3 \times 10^{36}$
erg s$^{-1}$, typical of type I outbursts.  With this value of the
distance and Galactic coordinates $l=87.12$, $b=-0.7$, \sax\  agrees with
a location within the Perseus spiral arm (Russeil \cite{rus03}).

Before our optical observations the ascription of \sax\ to the group of
Be/X-ray binaries was circumstantial and based on X-ray data only.  Our
spectroscopic observations in the spectral classification region exposed a
B0V star with a remarkable richness in spectral variability.  We conclude
that \sax\ belongs to the group of Be/X-ray binaries.

\begin{figure}
\resizebox{\hsize}{!}{\includegraphics{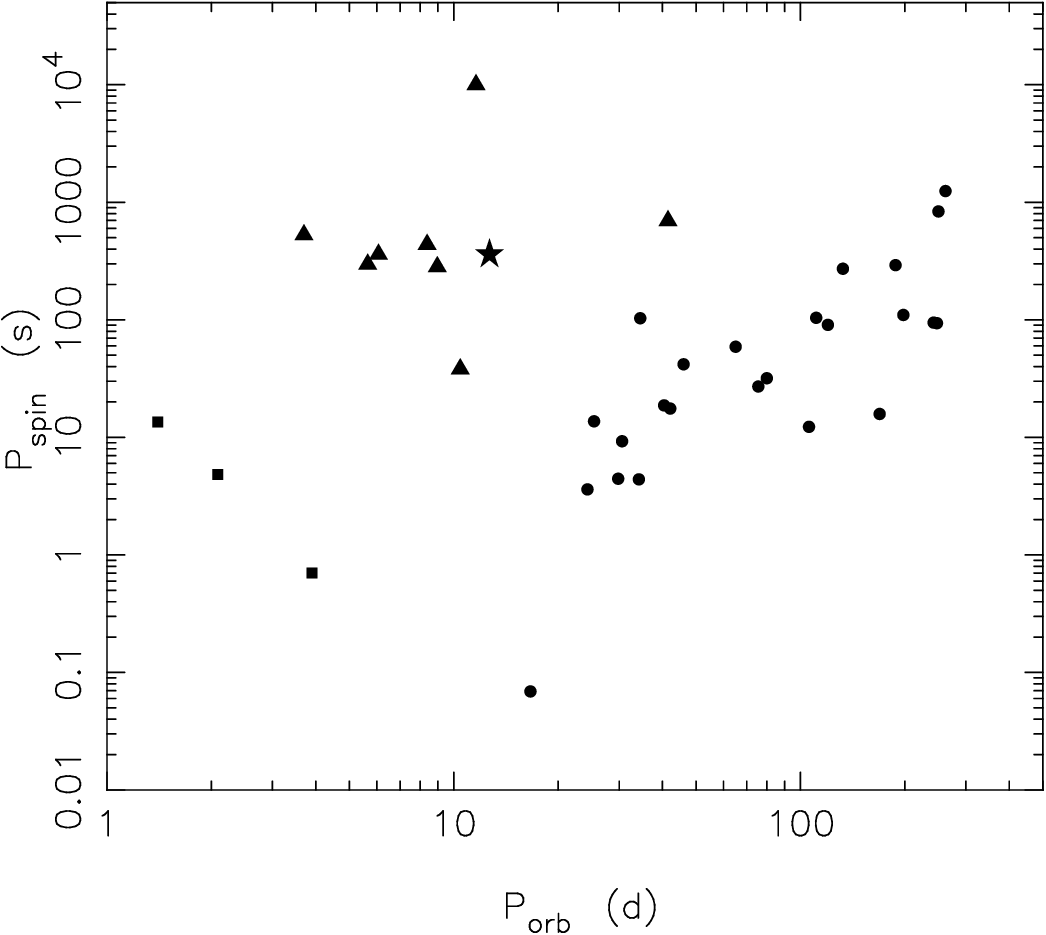}}
\caption[]{$P_{\rm spin}-P_{\rm orb}$ diagram. The star symbol marks the 
position of \sax. Circles represent Be/X-ray binaries, triangles identify
wind-fed supergiants and squares refer to disc-fed (Roche-lobe filling) 
systems.}
\label{pspo}
\end{figure}

\begin{table}
\begin{center}
\caption{Comparison of times of emission-to-absorption reversal of \sax\
with two other Be/X-ray systems}
\label{loss}
\begin{tabular}{cc|cc|cc}
\hline \hline \noalign{\smallskip}
\multicolumn{2}{c}{\sax$^a$}	&\multicolumn{2}{c}{4U\,0115+63$^b$}	&\multicolumn{2}{c}{X Per$^c$}  \\
MJD	&EW(H$\alpha$)  &MJD	&EW(H$\alpha$)  &MJD	&EW(H$\alpha$) 	\\
	&($\AA$)	&	&($\AA$)	&	&($\AA)$ \\
\noalign{\smallskip} \hline \noalign{\smallskip}
52853	&--2.20	&50481	&--3.6	&47925	&--1.3	\\
52869	&--1.07	&50647	&+1.3	&47950	&--1.14	\\
52897	&+2.32	&50767	&--0.4	&48137	&+1.6	\\
	&	&	&	&48194	&+2.7	\\
\noalign{\smallskip} \hline
\multicolumn{6}{l}{$a$: This work}	\\
\multicolumn{6}{l}{$b$: Negueruela et al. (\cite{neg01})}	\\
\multicolumn{6}{l}{$c$: Clark et al. (\cite{cla01})}
\end{tabular}
\end{center}
\end{table}

\subsection{Disc truncation}

Figure~\ref{ha} shows the profile of the H$\alpha$ line in \sax. The
corresponding equivalent widths are given in Table~\ref{ewha}.  The
reversal of the H$\alpha$ line from emission to absorption is one of the
most spectacular types of variability among Be stars. It is interpreted as
the loss of the circumstellar disc. Disc-loss episodes have been seen in a
number of Be/X-ray binaries: X-per (Roche et~al. \cite{roc93}), A0\,535+26
(Negueruela et al. \cite{neg00}), RX\,J0812.4--3114  (Reig et~al.
\cite{reig01}), 4U\,0115+63 (Negueruela et~al. \cite{neg01}). During these
episodes the strength of the H$\alpha$ line and the photometric magnitudes
and colours gradually (on time scales of months to years) fade away until
the disc is lost. In \sax, these changes seem to be extremely fast. 
Although in order to fully appreciate the time scales for disc loss one
should look at the long-term H$\alpha$ variability, it is instructive to
compare the time scales of the last instances of the
emission-to-absorption reversal of \sax\ with those of 4U\,0115+63 and X
Per (Negueruela et~al. \cite{neg01}; Clark et al. \cite{cla01}), as shown
in Table~\ref{loss}. For RX\,J0812.4--3114 and A0\,535+26 the observation
are too sparse and do not allow us to constrain the time scales. For a
similar change in the equivalent width of the H$\alpha$ line, \sax\
requires approximately 3 and 5 times less time than 4U\,0115+63 and
X\,Per, respectively. 

Given the small orbit and moderate eccentricity of the system disc
truncation by the neutron star must have an important effect on the
structure of the disc. According to the viscous decretion disc model, disc
truncation is  more effective in systems with low eccentricity ($e < 0.2$)
than in systems with high eccentricity ($e > 0.6$) owing to their wider
gap between the outer disc radius and the critical Roche lobe radius
(Okazaki \& Negueruela \cite{okaneg01}; Negueruela \& Okazaki
\cite{negoka01}). \sax, has the smallest orbit among accretion-powered
Be/X-ray binaries and is moderately eccentric ($e\sim 0.4$).  The
combination of these two factors results in rather efficient disc
truncation but also in small gaps that allow transfer of matter. Since
mass transfer in Be/X-ray binaries occurs via the $L_1$ point of the
critical Roche lobes, some amount of matter will be  captured by the
neutron star, hence producing outbursts. 

Assuming typical values of the mass and radius of a B0V star ($M_*=20
\msun$ and $R_*=8 \rsun$, Vacca et al. \cite{vac96}) and taking $e=0.4$,
the periastron distance is estimated in $\sim$ 38 $\rsun$ or $\sim$5
$R_*$. An estimate of the  critical lobe radius yields virtually the same
values. Thus, the neutron star acts very efficiently in removing angular
momentum from the disc and preventing its growth. Another consequence for
systems with small orbits is the absence of type II outburst --- major
increases in intensity, $L_{\rm X} \simmore 10^{37}$ erg s$^{-1}$, that
last for weeks or months and are usually accompanied by strong spin-up
episodes of the neutron star --- since the star does not have the
opportunity of developing an extended and steady disc. 

On the other hand, it is possible that a proper circumstellar disc never
forms. A scenario like that seen in  $\mu$ Cen (Baade et al. \cite{baa88};
Hanuschik et al. \cite{han93}), where fast (2-5 days) ejections of
photospheric material are lifted into bound orbits forming a short-lived
($\simless$ 100 days) quasi-stable Keplerian disc might be applicable to
\sax\ as well. H$\alpha$ outbursts of this type are never very intense,
the equivalent width does not go over a few \AA\ and are accompanied by
$V/R$ variability. Disc radii in such scenario do not go over 4-6 stellar
radii (Hanuschik et al. \cite{han93}).  The long periods of X-ray activity
in 1997 and 1999/2000 would, however, argue against this scenario.

\subsection{The spin period--orbital period correlation}

The available data clearly demonstrate that \sax\ is a Be/X-ray binary:
{\em i)} in the X-ray band it appears as a moderately eccentric transient
system that shows type I outbursts,  {\em ii)} it shows X-ray pulsations and the
X-ray spectral parameters are consistent with those found in other
Be/X-ray systems, {\em iii)} the optical/IR counterpart is a B0Ve star
that shows highly variable H$\alpha$ emission. Still, if the 12.7 day
modulation indeed represents the orbital period, then \sax\ is the
accretion-powered Be/X-ray binary with the shortest orbital period.


It is a well known fact that the two types of high-mass X-ray binaries
(HMXB), Be/X, and supergiant systems (SXRB) occupy well-defined positions
in the spin period versus orbital period diagram (Corbet \cite{cor86}).
Figure~\ref{pspo} displays an updated version of Corbet's diagram. \sax\
deviates from the correlation of Be/X systems and falls in the region of
wind-fed supergiant X-ray binaries. The correlation in Be/X systems is
explained in terms of the equilibrium period, defined as the period at
which the outer edge of the magnetosphere rotates with the Keplerian
velocity (Davidson \& Ostriker \cite{dav73}; Stella et al. \cite{ste86};
Waters \& van Kerkwijk \cite{wat89}).

According to Waters \& van Kerwijk (\cite{wat89}), wind-fed SXRB rotate at the
value of the equilibrium period that they had when they were main-sequence
stars. The subsequent wind-driven accretion torques were not strong enough
to spin up the neutron star after the spin-down that followed the
supernova explosion. In Be/X-ray binaries, the slow spin-down that
followed the supernova explosion was compensated by rapid spin-up episodes
due to the much more efficient ways of transfer angular momentum to the
neutron star through the much higher density equatorial winds. If this
interpretation is correct, then a Be/X-ray system going through a long
period of quiescence can spin down to the equilibrium period that
corresponds to the wind-driven accretion mechanism. In other words, in a
quiescent Be/X-ray binary (presumably due to the lack of circumstellar
matter) accretion takes place through the normal wind of an early-type
star. Therefore, the neutron star would rotate at a rate more appropriate
of a SXRB than of an active Be/X.

The spin-up rate of \sax\ during the first $\sim$ 30 days of the 1999
outburst was $2.5 \times 10^{-13}$ Hz s$^{-1}$ (Baykal et al.
\cite{bay00}). This value is comparable to those observed in other
Be/X-ray binaries during type I outbursts, where typical spin-up rates
$\simless \, 5 \times 10^{-12}$ Hz s$^{-1}$ have been reported (Wilson et
al. \cite{wil97}, \cite{wil02}; Finger et al. \cite{fin96}, \cite{fin99}).
However, the episodic spin-up phases in \sax\ are not capable to exert
strong enough torques on the neutron star to make it rotate at the
equilibrium period. A more efficient way to spin up the neutron star would
be through type II outbursts, for which typical values are $\simmore \, 8
\times 10^{-12}$ Hz s$^{-1}$ (Bildsten et al. \cite{bil97} and references
therein). These kind of outbursts, however, require high mass accretion
rates that can only be achieved with a large and extended or dense disc.
As explained above, due to the disc truncation \sax\ can never develop
such a disc.

In this respect, it is instructive to compared \sax\ with the Be/X-ray
system with the second shortest orbital period, A0538--66. In contrast to
\sax\, a spin period of only 66 ms has been reported  for A0538--66
(Skinner et al. \cite{ski82}), which would be indicative of large amounts
of angular momentum transfer. In the early 80's A0538-66 experienced  a
series of very luminous (super-Eddington) outbursts regularly separated by
16.7 d, which was interpreted as the orbital period. In the standard model
for this source (Charles et al. \cite{cha83}), it is believed that the
eccentricity of the neutron star's orbit is very high, so that the neutron
star comes very close to the surface of the B-type star at periastron,
sometimes inducing moderate Roche-lobe overflow. In addition,  A0538--66
is extremely active in the optical band (Alcock et al. \cite{alc01}) and high
accretion rates must be sustained. Hence, we can suspect that spin-up
episodes must be occurring frequently in A0538--66.

\section{Conclusions}

We have identified the optical counterpart to the X-ray transient \sax.
The optical spectroscopic observations revealed a B0Ve star displaying fast
spectral changes. Due to the short orbital period and moderate
eccentricity, the neutron star truncates the Be star's disc at a small
radius and prevents the development of an extended and steady disc.  The
impossibility to develop a dense disc implies that most of the time the
density and velocity profiles of the stellar wind, i.e, of the accreted
material, are those of a normal B star rather than the much slower and
denser wind of a Be star. The result is a slow neutron star that rotates
far away from its equilibrium period. The equatorial disc formed under
these conditions is short-lived and perturbed, having density gradients
that lead to the formation of asymmetries that translate into V/R
variability. \sax\ is an excellent example of how the neutron star affects
the physical properties of the Be star companion.

\begin{acknowledgements}

P.~R. and I.~N. are  researchers of the programme {\em Ram\'on y Cajal}
funded by the Spanish Ministery of Science and Technology and the
Universities of Valencia and Alicante, under grants ESP2002-04124-C03-01
and ESP2002-04124-C03-03, respectively. P.~R. acknowledges partial support
by the Agencia Velenciana de Ciencia y Tecnolog\'{\i}a under grant
CTESPP/2003/002. Skinakas Observatory is a collaborative project of the
University of Crete, the Foundation for Research and Technology-Hellas and
the Max-Planck-Institut f\"ur Extraterrestrische Physik. Some of the WHT
spectra were obtained as part of the ING service programme.

\end{acknowledgements}

\end{document}